\documentclass[aps,prl,reprint,superscriptaddress,floatfix,amsmath,nofootinbib]{revtex4-1}
\usepackage{graphicx,epstopdf,natbib,color}

\newcommand{\AsAs}{\aap}
\newcommand{\npa}{\nphysa}
\renewcommand{\prc}{Phys. Rev. C}

\renewcommand{\npa}{Nucl. Phys. A}
\renewcommand{\prl}{Phys. Rev. Lett.}
\newcommand{\aap}{Astron. Astrophys.}
\newcommand{\mnras}{Mon. Not. R. Astron Soc.}
\newcommand{\nima}{Nucl. Instrum. Meth. A}
\newcommand{\nimb}{Nucl. Instrum. Meth. B}
\newcommand{\plb}{Phys. Lett. B}

\newcommand{\nds}{Nucl. Data Sheets}
\newcommand{\cpc}{Chinese Phys. C}
\begin{document}

\title{$\beta$-decay of $^{61}$V and its Role in Cooling Accreted Neutron Star Crusts}

\author{W.-J.~Ong}
\email[]{ong10@llnl.gov}
\affiliation{Nuclear and Chemical Sciences Division, Lawrence Livermore National Laboratory, Livermore, CA 94550, USA}
\affiliation{Department of Physics and Astronomy, Michigan State University, East Lansing, MI 48824, USA}
\affiliation{National Superconducting Cyclotron Laboratory, East Lansing, MI 48824, USA}

\author{E.~F.~Brown}
\affiliation{Department of Physics and Astronomy, Michigan State University, East Lansing, MI 48824, USA}
\affiliation{National Superconducting Cyclotron Laboratory, East Lansing, MI 48824, USA}
\affiliation{Joint Institute for Nuclear Astrophysics \textrm{--} Center for the Evolution of the Elements, Michigan State University, East Lansing, MI 48824, USA}
\affiliation{Department of Computational Mathematics, Science, and Engineering, Michigan State University, East Lansing, MI 48824, USA}

\author{J.~Browne}
\affiliation{Department of Physics and Astronomy, Michigan State University, East Lansing, MI 48824, USA}
\affiliation{National Superconducting Cyclotron Laboratory, East Lansing, MI 48824, USA}

\author{S.~Ahn}
\affiliation{Cylotron Institute, Texas A$\&$M University, College Station, TX 77843, USA}
\affiliation{Joint Institute for Nuclear Astrophysics \textrm{--} Center for the Evolution of the Elements, Michigan State University, East Lansing, MI 48824, USA}

\author{K.~Childers}
\affiliation{Department of Chemistry, Michigan State University, East Lansing, MI 48824, USA}
\affiliation{National Superconducting Cyclotron Laboratory, East Lansing, MI 48824, USA}

\author{B.~P.~Crider}
\affiliation{Department of Physics and Astronomy, Mississippi State University, Mississippi State, MS 39762, USA}

\author{A.~C.~Dombos}
\affiliation{Department of Physics and Astronomy, Michigan State University, East Lansing, MI 48824, USA}
\affiliation{National Superconducting Cyclotron Laboratory, East Lansing, MI 48824, USA}
\affiliation{Joint Institute for Nuclear Astrophysics \textrm{--} Center for the Evolution of the Elements, Michigan State University, East Lansing, MI 48824, USA}

\author{S.~S.~Gupta} 
\affiliation{Indian Institute of Technology Ropar, Nangal Road, Rupnagar (Ropar), Punjab 140 001, India}

\author{G.~W.~Hitt}
\affiliation{Department of Physics and Engineering Science, Coastal Carolina University, Conway, SC 29528, USA}

\author{C.~Langer}
\affiliation{Institute for Applied Physics, Goethe-University Frankfurt a. M., Frankfurt am Main 60438, Germany}

\author{R.~Lewis}
\affiliation{Department of Chemistry, Michigan State University, East Lansing, MI 48824, USA}
\affiliation{National Superconducting Cyclotron Laboratory, East Lansing, MI 48824, USA}

\author{S.~N.~Liddick}
\affiliation{Department of Chemistry, Michigan State University, East Lansing, MI 48824, USA}
\affiliation{National Superconducting Cyclotron Laboratory, East Lansing, MI 48824, USA}

\author{S.~Lyons}
\affiliation{National Superconducting Cyclotron Laboratory, East Lansing, MI 48824, USA}
\affiliation{Joint Institute for Nuclear Astrophysics \textrm{--} Center for the Evolution of the Elements, Michigan State University, East Lansing, MI 48824, USA}

\author{Z.~Meisel}
\affiliation{Department of Physics and Astronomy, Ohio Univeristy, Athens, OH 45701, USA}
\affiliation{Joint Institute for Nuclear Astrophysics \textrm{--} Center for the Evolution of the Elements, Michigan State University, East Lansing, MI 48824, USA}

\author{P.~M{\"o}ller}
\affiliation{Theoretical Division, Los Alamos National Laboratory, Los Alamos, NM 87545, USA}
\affiliation{Joint Institute for Nuclear Astrophysics \textrm{--} Center for the Evolution of the Elements, Michigan State University, East Lansing, MI 48824, USA}

\author{F.~Montes}
\affiliation{National Superconducting Cyclotron Laboratory, East Lansing, MI 48824, USA}
\affiliation{Joint Institute for Nuclear Astrophysics \textrm{--} Center for the Evolution of the Elements, Michigan State University, East Lansing, MI 48824, USA}

\author{F.~Naqvi}
\affiliation{National Superconducting Cyclotron Laboratory, East Lansing, MI 48824, USA}
\affiliation{Joint Institute for Nuclear Astrophysics \textrm{--} Center for the Evolution of the Elements, Michigan State University, East Lansing, MI 48824, USA}
\affiliation{Department of Physics \& Astrophysics, University of Delhi, Delhi 110007, India}

\author{J.~Pereira}
\affiliation{National Superconducting Cyclotron Laboratory, East Lansing, MI 48824, USA}
\affiliation{Joint Institute for Nuclear Astrophysics \textrm{--} Center for the Evolution of the Elements, Michigan State University, East Lansing, MI 48824, USA}

\author{C.~Prokop}
\affiliation{Los Alamos National Laboratory, Los Alamos, NM 87545, USA}

\author{D.~Richman}
\affiliation{Department of Physics and Astronomy, Michigan State University, East Lansing, MI 48824, USA}
\affiliation{Los Alamos National Laboratory, Los Alamos, NM 87545, USA}

\author{H.~Schatz}
\affiliation{Department of Physics and Astronomy, Michigan State University, East Lansing, MI 48824, USA}
\affiliation{National Superconducting Cyclotron Laboratory, East Lansing, MI 48824, USA}
\affiliation{Joint Institute for Nuclear Astrophysics \textrm{--} Center for the Evolution of the Elements, Michigan State University, East Lansing, MI 48824, USA}

\author{K.~Schmidt}
\thanks{Present Address: Institute of Radiation Physics, Helmholtz-Zentrum Dresden-Rossendorf, Dresden 01328, Germany}
\affiliation{National Superconducting Cyclotron Laboratory, East Lansing, MI 48824, USA}
\affiliation{Joint Institute for Nuclear Astrophysics \textrm{--} Center for the Evolution of the Elements, Michigan State University, East Lansing, MI 48824, USA}

\author{A.~Spyrou}
\affiliation{Department of Physics and Astronomy, Michigan State University, East Lansing, MI 48824, USA}
\affiliation{National Superconducting Cyclotron Laboratory, East Lansing, MI 48824, USA}
\affiliation{Joint Institute for Nuclear Astrophysics \textrm{--} Center for the Evolution of the Elements, Michigan State University, East Lansing, MI 48824, USA}
\date{\today}

\begin{abstract}

%motivation
The interpretation of observations of cooling neutron star crusts in quasi-persistent X-ray transients is affected by predictions of the strength of neutrino cooling via crust Urca processes. The strength of crust Urca neutrino cooling depends sensitively on the electron-capture and $\beta$-decay ground-state to ground-state transition strengths of neutron-rich rare isotopes. Nuclei with mass number $A=61$ are predicted to be among the most abundant in accreted crusts, and the last remaining experimentally undetermined ground-state to ground-state transition strength was the $\beta$-decay of $^{61}$V. 
%result 
This work reports the first experimental determination of this transition strength, a ground-state branching of 8.1$^{+2.2}_{-2.0} \%$, corresponding to a log $ft$ value of 5.5$^{+0.2}_{-0.2}$. 
%method
This result was achieved through the measurement of the $\beta$-delayed $\gamma$ rays using the total absorption spectrometer SuN and the measurement of the $\beta$-delayed neutron branch using the neutron long counter system NERO at the National Superconducting Cyclotron Laboratory at Michigan State University. This method helps to mitigate the impact of the Pandemonium effect in extremely neutron-rich nuclei on experimental results.  
%impact and conclusion
The result implies that $A=61$ nuclei do not provide the strongest cooling in accreted neutron star crusts as expected by some predictions, but that their cooling is still larger compared to most other mass numbers. Only nuclei with mass numbers 31, 33, and 55 are predicted to be cooling more strongly. However, the theoretical predictions for the transition strengths of these nuclei are not consistently accurate enough to draw conclusions on crust cooling. With the experimental approach developed in this work all relevant transitions are within reach to be studied in the future. 
\end{abstract}

\maketitle

%%%%Introduction%%%%
%%%%Introduction%%%%
%%%%Introduction%%%%

X-ray observations of the cooling of transiently accreting neutron stars provide insights into the properties of the star. In quasi-persistent systems where accretion turns off for years, long-term observations reveal the thermal profile of the crust, which probes heat capacity and heat-transport properties of dense matter (see \cite{Meisel_Review2018} for a recent review). The crust, an outer layer where the nuclei are arranged in a lattice, is built up from the hot ashes of thermonuclear X-ray bursts that occur on the surface of the neutron star during the accretion phase \cite{Galloway_Review2017,Schatz2006a}. These ashes are incorporated into the neutron star crust by ongoing accretion and converted into increasingly neutron-rich species through electron captures that occur when the Fermi energy of the degenerate electrons exceeds the electron-capture energy thresholds \cite{bisnovatyui1979,Sato1979,Haensel1990,Gupta2007,Gupta2008,Haensel2008,Lau2018}.  It has been shown that under realistic crust conditions at non-zero temperatures, the thermal elevation of electrons above the Fermi surface allows in some cases the reverse $\beta$-decay reactions to occur in addition to the electron-capture reactions \cite{Schatz2014}. The resulting cycle of alternating electron captures and $\beta$ decays between the same pair of nuclei can lead to rapid neutrino cooling. If efficient, such a crust Urca process can impact the cooling behavior of the neutron star and has to be taken into account when interpreting X-ray observations of cooling neutron stars \cite{Deibel2016,Meisel2017}. 
 
Within an electron-capture sequence along an isobaric mass chain, strong Urca cooling occurs when there are strong ground-state to ground-state electron-capture and $\beta$-decay transitions, and when the subsequent electron capture to the $A, Z-2$ nucleus is blocked \cite{Schatz2014}. The abundance of nuclei with a certain mass number in the nuclear ashes of the X-ray bursts determines which mass chains are populated in the neutron star crust of a given system.  Significant cooling is expected from Urca pairs in the $A=31,33$ mass chains, due in part to their significant abundance in rp-process ashes \cite{Cyburt2016}; however, the composition of the burst ashes in this mass region is uncertain as it depends on freezeout conditions and residual helium burning \cite{Woosley2004a}. An important question addressed here is whether there is any significant cooling from the ashes of the rp-process, which predominantly produces nuclei in the $A=56\textrm{--}72$ mass range \cite{Woosley2004a,Jose2010, Cyburt2016}. The theoretical model to predict electron capture and $\beta$-decay transition strengths used in current crust models is the  QRPA-fY  \cite{Moller1990,Moller1997} owing to its ability to make consistent predictions for all relevant nuclei \cite{MOLLER2016, MOLLER2019}. QRPA-fY predicts $A=56$ to be the strongest cooling isobaric chain, but recent experimental and theoretical results have shown that this mass chain does not in fact cool at all \cite{Meisel2015}. Based on the composition of the rp-process ashes, this leaves the odd $A$ chains $A=55,57,59,61,63,65$ (the most abundant odd-A ashes) as predicted candidates for Urca cooling transitions. In order to ascertain whether the most common accreted neutron star crusts from mixed H/He bursts exhibit significant crust Urca cooling, and to quantify the neutrino cooling rates, it is important to experimentally constrain the ground-state to ground-state transitions in these mass chains. We present here an experimental approach to provide such constraints, and apply it to the $A=61$ mass chain for the first time. 

Of all the relevant odd-A nuclides in X-ray burst ashes, $A=61$ nuclei are second most abundant (after $A=65$). It has also been shown that within the uncertainty of the $^{61}$Ga(p,$\gamma$)$^{62}$Ge reaction rate in X-ray burst models, $A=61$ could even be the most dominant constituent in the rp-process ash \cite{Cyburt2016}. Of the four relevant electron-capture transitions in this mass chain that are located in the outer crust, experiments in the $\beta$ decay direction have established ground-state to ground-state transitions for the first three, with strengths of log $ft_{\beta}$($^{61}$Fe) $\geq$ 7.1 \cite{Ehrlich1967}, log $ft_{\beta}$($^{61}$Mn) = 5.02(3) \cite{Radulov2013}, and log $ft_{\beta}$($^{61}$Cr) = 5.1(2) \cite{Crawford2009}.  Given these transition strengths and the smaller electron-capture $Q$-values, Urca cooling in the $A=61$ chain occurs, but is relatively weak. However, for the fourth electron capture from $^{61}$Cr to $^{61}$V, an allowed ground-state to ground-state transition is also possible when considering selection rules and the estimated ground-state spins for $^{61}$Cr ($5/2^-$) and $^{61}$V ($3/2^-$).  QRPA-fY theory does not predict a ground-state to ground-state electron-capture transition (the lowest-lying transition from the ground state is predicted to a state with 3 MeV excitation energy in $^{61}$V). However, QRPA-fY does predict a strong transition (log $ft$=4.35) in the $\beta$-decay direction from $^{61}$V to a low-lying state with an excitation energy of just 10~keV in $^{61}$Cr. Within the theoretical uncertainties this excitation energy would be consistent with the ground state. If indeed there were a strong ground-state to ground-state transition with log $ft$=4.35,  $A=61$ would become the most important Urca cooling chain for rp-process ashes, even without a larger $A=61$ abundance from a lower than predicted  $^{61}$Ga(p,$\gamma$)$^{62}$Ge reaction rate in X-ray bursts. In the $\beta$-decay direction, such a strong transition would still be compatible with the experimentally determined 48.3~ms half-life \cite{Zuber20151} within the relatively large Q-value uncertainties of 12.0$\pm$0.9 MeV \cite{AME16} but would require a $>65\%$ ground-state to ground-state branching for the decay. 

Previous $\beta$-delayed $\gamma$-spectroscopy studies of $^{61}$V using an array of high-purity germanium detectors deduced an upper limit for the $\beta$-decay branch to the ground state of $^{61}$Cr of 40$\%$ \cite{Suchyta2014}. This corresponds to a lower limit on the log $ft$ value of 4.6, still a very strong Urca cycle. This limit was obtained from identifying transitions to 12 excited states and determining the corresponding $\beta$ decay feeding intensities.  This level scheme is incomplete, as only states up to 2.26 MeV were identified, while the $\beta$-decay Q-value is $\approx$12 MeV, and it is likely that the deduced $\beta$-decay feeding intensities reported were subject to the Pandemonium effect \cite{HARDY1977307}. Here, we report the first determination of the $\beta$ decay branch of $^{61}$V to the ground state and use an experimental approach that combines use of the total absorption $\gamma$-spectrometer SuN \cite{SIMON201316} and the neutron detector NERO \cite{Pereira2010}. With its high summing efficiency, SuN is capable of detecting even very weak $\gamma$-emitting transitions in $^{61}$Cr fed by the $\beta$ decay of $^{61}$V. This avoids the Pandemonium effect and enables accurate determination of all feeding intensities. NERO was used to determine the total $\beta$-delayed neutron emission branch, including the transition to the ground state of $^{60}$Cr that cannot be determined through $\gamma$-ray detection. The ground-state to ground-state branch is extracted from the total number of decays by accounting for the $\beta$-delayed $\gamma$ branchings to all exited states above and below the neutron separation energy in $^{61}$Cr and the $\beta$-delayed neutron branch to $^{60}$Cr measured with NERO.

%%%%Methods and Results%%%%
%%%%Methods and Results%%%%
%%%%Methods and Results%%%%
The experiment was performed at the National Superconducting Cyclotron Laboratory (NSCL) at Michigan State University. $^{61}$V was produced as part of a mixed secondary beam (36$\%$ $^{61}$V) by impinging a $^{82}$Se primary beam (140 MeV/u, 35 pnA) on a 352 mg/cm$^2$ Be target and purifying the ensuing fragment beam with the A1900 fragment separator \cite{morrissey2003commissioning}. The secondary beam was transported to the experimental end station where it was implanted into a double-sided Si strip detector (DSSD) at a total rate of $\approx$80 pps. An Si PIN detector upstream of the DSSD was used to characterize incoming beam particles by recording energy loss and the time of flight from the A1900 scintillator.

The experiment was carried out in two parts: In the first part $\approx$160,000 $^{61}$V ions were implanted into a DSSD that was part of the NSCL Beta Counting System \cite{prisciandaro2003beta} located in the center of the NERO neutron detector to detect $\beta$-delayed neutrons in coincidence with $\beta$-particles detected by the DSSD following an ion implantation. In the second part of the experiment, approximately one million $^{61}$V ions were implanted into a mini DSSD located at the target position of the SuN $\gamma$-ray detector to detect $\beta$-delayed $\gamma$ rays in coincidence with $\beta$ particles. The measured half-life of $^{61}$V was 48$\pm$1~ms, in agreement with previously measured values \cite{sorlin1999beta, Sorlin2003, Gaudefroy2005, Daugas2011,  NSR2002MAZN, Suchyta2014}. 

A branching for $\beta$-delayed neutron emission $P_\textrm{n}=$14.5 $\pm$ 2.0 $\%$ was determined from the number of neutrons detected in delayed coincidence with an implantation of a $^{61}$V beam particle. This is consistent with $P_\textrm{n}$ $>$ 12 $\%$ independently determined from the SuN data, which contains peaks from the $\gamma$ decay of the $\beta$-delayed neutron daughter $^{60}$Cr.  

\begin{figure}
\centering
\includegraphics[trim=2.6cm 2.2cm 2cm 0,clip,width=0.48\textwidth]{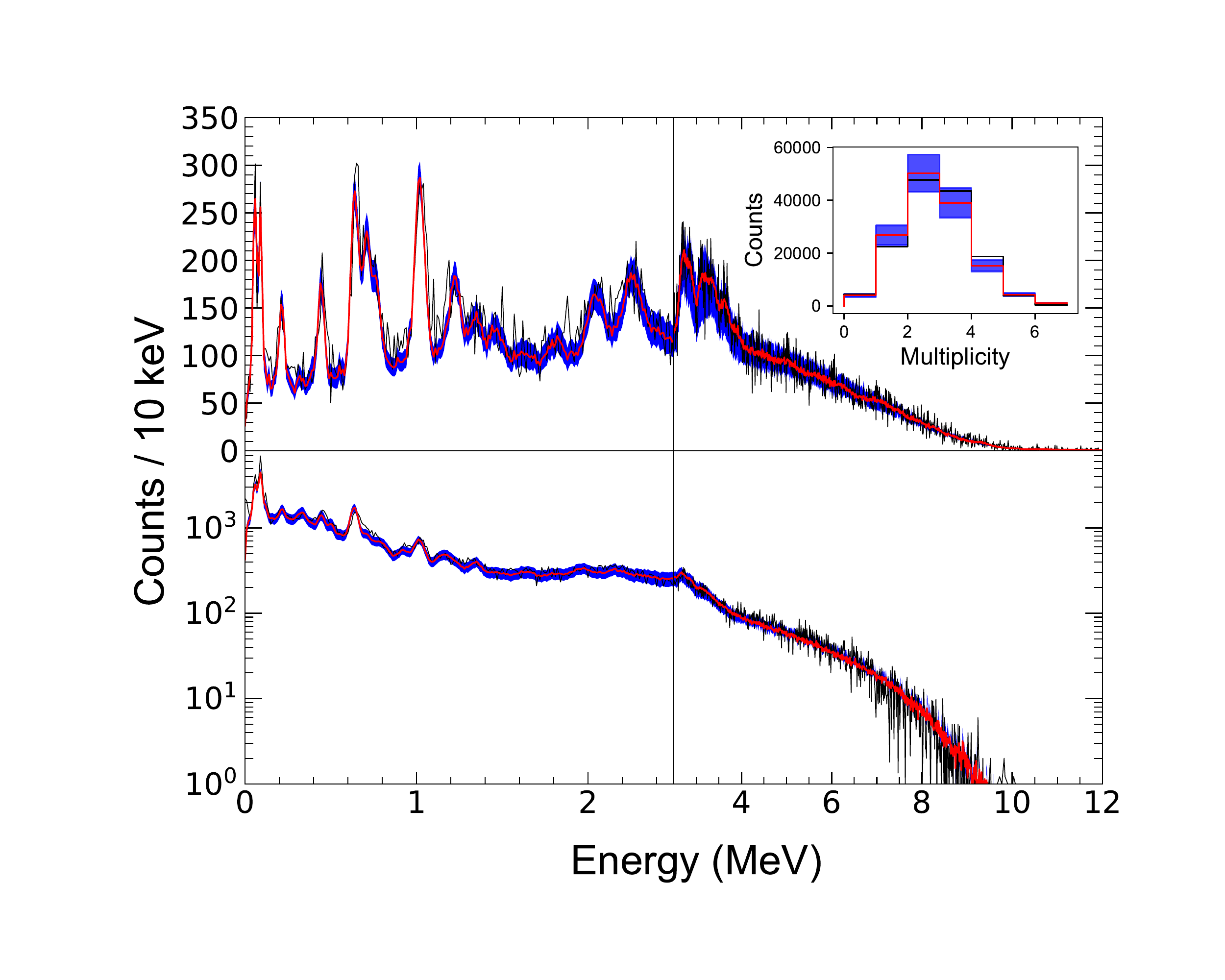}
\caption{Experimental (black) and fitted (red) total absorption spectrum (top), singles spectrum (bottom), and multiplicity distribution of the detector segment (inset). The 1-$\sigma$ error band of the fit is shown in blue. The high-energy tail in the total absorption spectrum is due to a combination of the high-energy $\beta$ particles from low-energy entry states entering SuN and $\gamma$-decay cascades from high-energy entry states.}
\label{TASspectrum}
\end{figure}

%%%%Analysis and Conclusion%%%%
%%%%Analysis and Conclusion%%%%
%%%%Analysis and Conclusion%%%%

For the SuN measurement, the energies deposited by all $\beta$-delayed $\gamma$ rays emitted in coincidence with each detected $\beta$ decay were summed together to create the total absorption spectrum shown in Fig.~\ref{TASspectrum}. The $\beta$-feeding intensity to each state in $^{61}$Cr is related to the background corrected area of the total absorption peak at the excitation energy of the state. The total absorption spectrum shows peaks for all previously identified states \cite{Suchyta2014}, but no additional isolated states were identified owing to the limited resolution of SuN and background from summing with $\beta$-particles. Nevertheless, the total absorption spectrum records the feeding of all additional states. To extract this information we followed the analysis described in \cite{dombos2016total, ong2018quantifying}. Templates of the total absorption spectra for $\beta$ decays to states in $^{61}$Cr and $\beta$-delayed neutron emission to states in $^{60}$Cr are generated using GEANT4 simulations. In addition to transitions to individual states, transitions to unknown levels above 2.26 MeV are treated as a quasi-continuum, with pseudolevels inserted every 50~keV up to 3.2 MeV, every 100~keV up to 3.95~MeV and every 200~keV up to the $Q$-value of 12.0 (0.9) MeV \cite{AME16}, following the resolution of the SuN detector. 
 
Because the $\gamma$-ray summing efficiency is less than 100\% there is a remaining dependency of the total absorption spectrum on the detailed $\gamma$-ray cascade emitted following the $\beta$-decay feeding of a state. For the individual states in the known level scheme, the cascade branchings were determined using spectra of individual segments of SuN. For the quasi-continuum states, the statistical code \textsc{Dicebox} \cite{Becvar1998434} was used, with input parameters for the nuclear level density and $\gamma$-ray strength functions taken from the RIPL-3 database \cite{CAPOTE2009}. The measured total absorption and $\gamma$-ray singles spectra were then simultaneously fitted as a linear combination of all templates. The $\beta$-decay transition strengths, including the $\beta$-delayed neutron emission feeding, are then the normalized fit coefficients. Fig.~\ref{TASspectrum} shows the good agreement of the template fit with the measured spectrum as well as the $\gamma$-ray multiplicity distribution. The $\chi^2$ values per degree of freedom are 1.26, 3.53 and 1.31 for the total absorption spectrum, the singles spectrum, and the combined data, respectively. The resulting $\beta$-feeding intensities are listed in Tab.~\ref{tab:feedings2}. The inferred feedings of low lying states from \cite{Suchyta2014, Zuber20151} (right column in Tab.~\ref{tab:feedings2}) are systematically larger because of missed transitions from higher-lying states. The significant feeding to states above the previously highest-known 2.26 MeV state inferred from our measurement also allows us to deduce a value, instead of an upper limit,  for the feeding of the ground state. 

To assess the error in the $\beta$-feeding intensities extracted from the fit, a Monte Carlo bootstrap study \cite{Booth1998} was performed with 100,000 drawn samples of synthetic data (greater than the validity sample number metric $n$ln($n$)$\approx$18,000 in this case) which were then re-fit to the same set of templates as the data using the same $\chi^2$ minimization procedure. The resultant distribution for each of the parameters was then used to determine the error in each parameter. Additionally, the impact of the uncertainty in the $\beta$-decay $Q$-value on the extracted $\beta$-feeding intensities was investigated by repeating the fit over the range of the uncertainty of the $Q$-value by varying the maximum energy of the pseudolevels included. The impact of $Q$-value uncertainties was found to be negligible compared to the error from the fit, as the feeding of the states with the highest excitation energy is relatively weak. Finally, the systematic uncertainty associated with uncertainties in the input parameters for \textsc{Dicebox} (such as the chosen models for the $\gamma$-ray strength function or level density), was assessed by repeating the $\chi^2$ minimization procedure with new pseudolevel templates generated when using different \textsc{Dicebox} inputs. About 50\% of the final uncertainties are due to this systematic error. 

Since the $ft$ values also depend on the Q-value through the Fermi integral, the Q-value uncertainty was also incorporated into the total uncertainties in the $ft$ values using Gaussian error propagation. The $Q$-value uncertainty is dominated by the mass uncertainty of $^{61}$V ($\Delta M = -30506.429$ $\pm$ 894.234 keV \cite{AME16}). If this uncertainty were reduced to 100~keV, reasonably achievable by Penning trap measurements at radioactive beam facilities\cite{Dilling2018} in the future, it would reduce the deduced log $ft$ uncertainties from $\pm$0.2 to $\pm$0.1.
 
\begin{table}
\centering
\caption{List of $\beta$-feeding intensities from this work (left column) to the identified states in the $^{61}$Cr excitation scheme from \cite{Suchyta2014}, and the corresponding apparent $\beta$ feedings deduced by \cite{Zuber20151} from \cite{Suchyta2014} (right column). The errors given here reflect both the statistical and systematic errors.}
\begin{tabular}{ccc}
\hline 
State (keV) & This work ($\%$) & Suchyta \textit{et al.} \cite{Suchyta2014, Zuber20151} ($\%$) \\
\hline \hline
\\[-0.7em]
0 & 8.1$^{+2.2}_{-2.0}$ & \textless 40 \\
\\[-0.7em]
65 & 2.9$^{+0.4}_{-0.4}$ & 6.0 (1.2) \\
\\[-0.7em]
97 & 2.1$^{+0.4}_{-0.4}$ & 10 (6) \\
\\[-0.7em]
224 & 2.0$^{+0.3}_{-0.3}$ & 7.1 (1.1)\\
\\[-0.7em]
402 & 0.17$^{+0.28}_{-0.17}$ & 1.7 (0.5) \\
\\[-0.7em]
451 & 3.0$^{+0.3}_{-0,4}$ & 5.7 (0.7) \\
\\[-0.7em]
564 & 0.26$^{+0.25}_{-0.21}$ & 1.5 (0.4)\\
\\[-0.7em]
632 & 0.0                                 & 1.7 (0.7) \\
\\[-0.7em]
716 & 4.4$^{+0.6}_{-0.6}$ & 6.1 (0.8)\\
\\[-0.7em]
774 & 3.0$^{+0.5}_{-0.5}$ & 3.2 (0.7) \\
\\[-0.7em]
1028 & 8.9$^{+0.6}_{-0.6}$ & 8.4 (0.8)\\
\\[-0.7em]
1233 & 4.5$^{+0.5}_{-0.5}$ & 1.4 (0.3)\\
\\[-0.7em]
2055 & 4.0$^{+0.4}_{-0.5}$ & 2.2 (0.7) \\
\\[-0.7em]
2262 & 5.8$^{+0.7}_{-0.7}$ & 2.5 (0.8) \\
\\[-0.7em]
Quasi-continuum & 36$^{+9}_{-9}$ & - \\
\\[-0.7em]
$P_{\textrm{n}}$ & 15$^{+3}_{-2}$ & - \\
\\[-0.7em]
\hline
\end{tabular}
\label{tab:feedings2}
\end{table}

\begin{figure}
\centering
\includegraphics[trim=2.5cm 1cm 1.8cm 0, clip, width=0.48\textwidth]{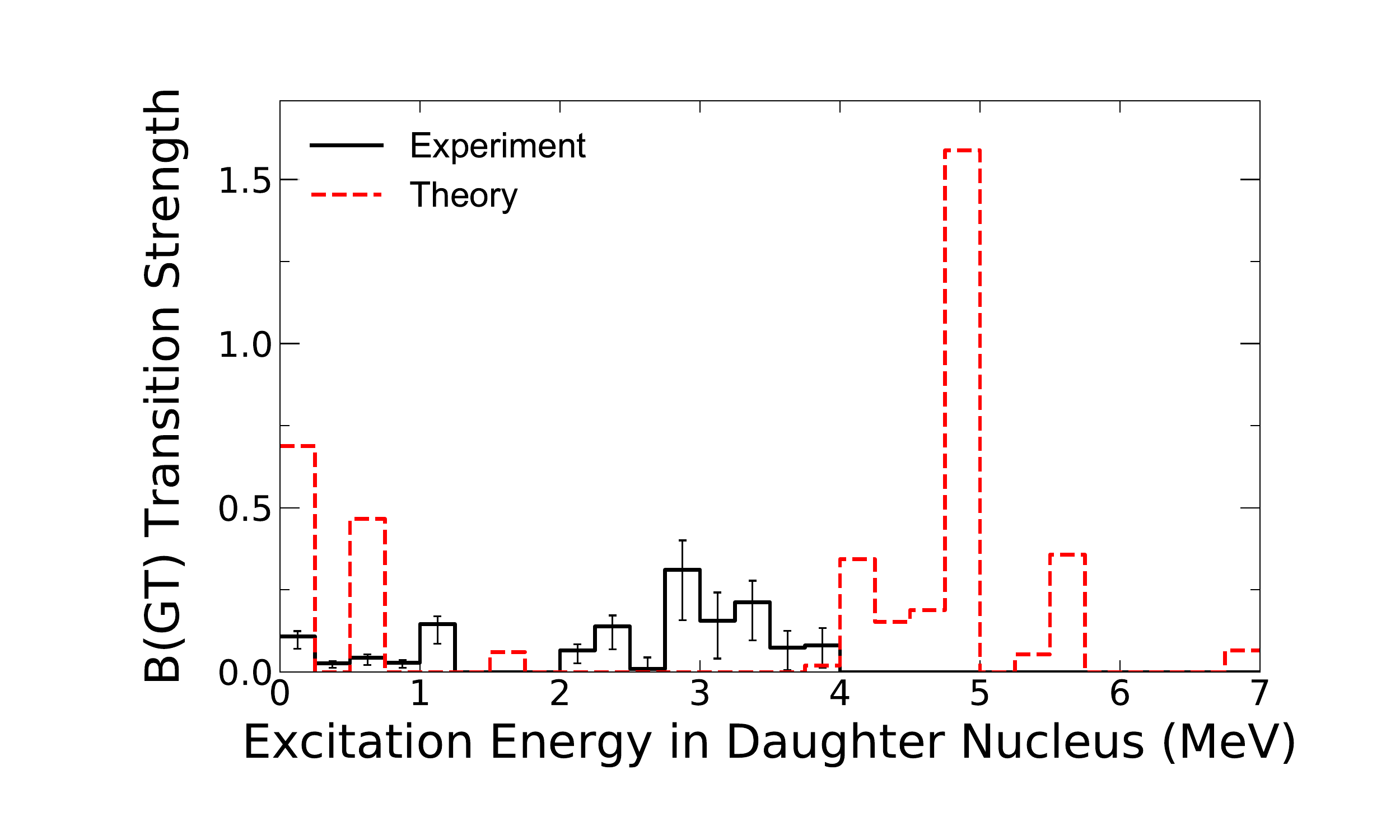}
\caption{B(GT) strength for the $\beta$-decay of $^{61}$V as function of excitation energy in the daughter nucleus, deduced from the $\gamma$-ray data from this work (solid black) and predicted by QRPA-fY (red, dashed). Note that $\gamma$-ray data can only provide the strength function up to the neutron separation energy of 3.9 MeV.}
\label{FigStrength}
\end{figure}

Fig.~\ref{FigStrength} compares the deduced $B($GT) strengths from our work (assuming that all transitions are allowed) below the $^{61}$Cr neutron separation energy $S_n=3.9 \pm 0.2$~MeV \cite{AME16} with the strength distribution predicted by QRPA-fY.  The calculations use the moderate oblate deformation of $\epsilon_2 \approx -0.1$ predicted by the FRDM for $^{61}$V \cite{FRDM95}. This would imply a $\nu$[321]3/2$^-$ ground state for $^{61}$Cr and a $\pi$[303]5/2$^-$ ground state for $^{61}$V (e.g. \cite{SORLIN1998205}). Previous work had tentatively assigned ground-state spins and parities of 5/2$^-$ for $^{61}$Cr and 3/2$^-$ for $^{61}$V and explained these with a moderate prolate deformation \cite{Suchyta2014}. Either scenario results in an allowed transition and is thus consistent with our data. 

QRPA-fY correctly predicts a transition to around the ground state. However, this transition is to an excited state at 10 keV in $^{61}$Cr and is therefore not included in the predicted electron-capture transitions on the $^{61}$Cr ground state. Current crust models that only consider ground-state electron captures \cite{Lau2018} therefore do not include a $^{61}$Cr--$^{61}$V Urca cooling pair. This shows the importance of including low-lying excited parent states in electron-capture transitions. Overall, theory predicts significant strength at low excitation energies in line with experimental results, though the measured strength is more spread out than predicted, resulting in less strength near the ground state. The strength function above the neutron separation energy is probed by the measured $P_n$-value. Here, the predicted value from QRPA-fY of 19\% is in good agreement with our measured value of 14.5 $\pm$ 2 $\%$. One reason for the more fragmented distribution of transition strengths seen in the experimental data is the contributions from different deformations. These contributions are not included in the QRPA-fY calculations, where it is assumed that the structure of the nucleus can be calculated at a precise single deformation. No standard, tested procedure to use superimposed states of different shape in the theory has yet been developed. This is consistent with previous work pointing out the importance of shape coexistence in describing nuclei in this region \cite{Lid11}. 

\begin{figure}
\centering
\includegraphics[trim=2.5cm 0 2cm 0,clip,width=0.48\textwidth]{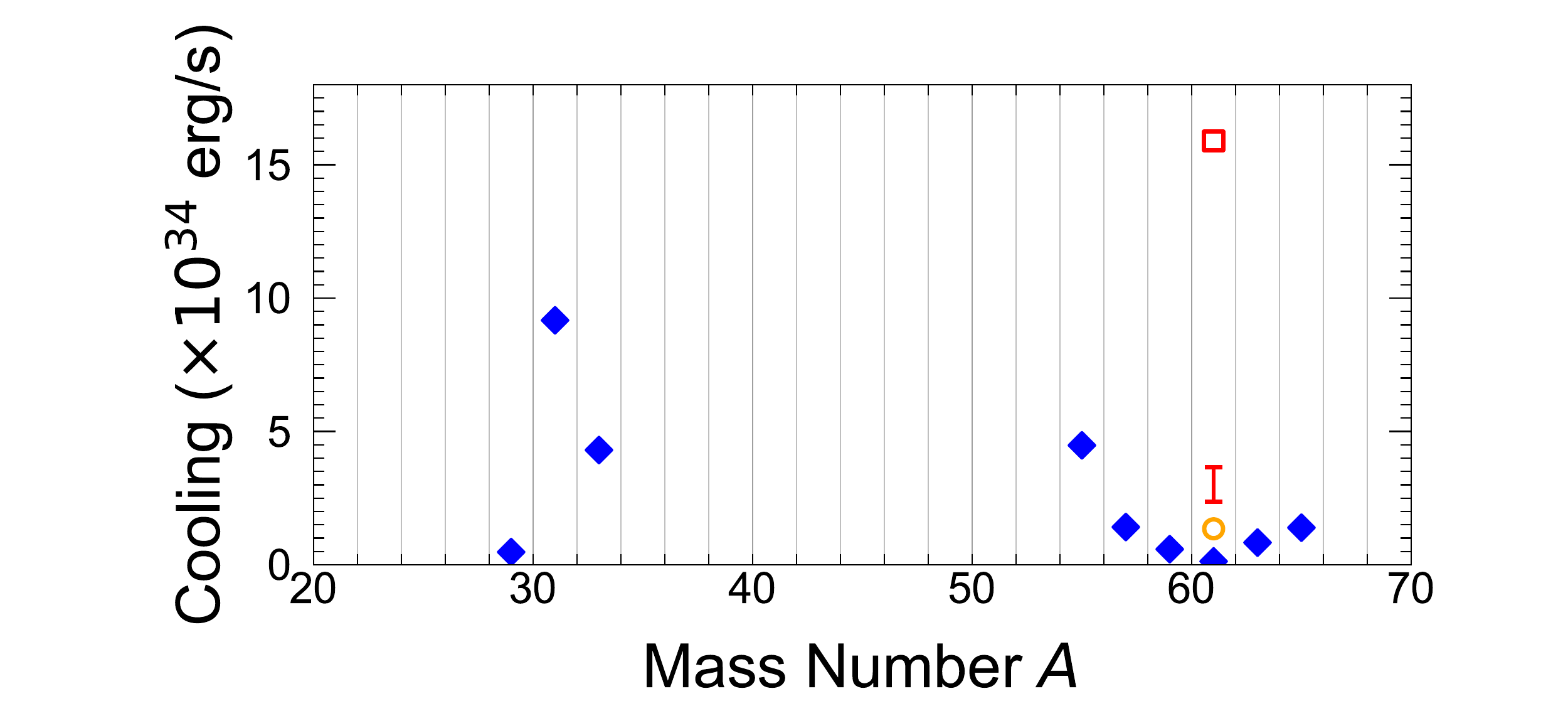}
\caption{The neutrino cooling contribution from individual mass chains for a neutron star crust made from rp-process ashes \cite{Cyburt2016}. Predictions based on QRPA-fY (blue filled diamonds) are shown for the strongest cooling chains. For $A=61$ we also show the QRPA-fY prediction when assuming the predicted 10 keV daughter state in $^{61}$Cr populated by the  $\beta$-decay of $^{61}$V is in fact the ground state (red open square, see text for more discussion). The $A=61$ cooling rates based on experimental transition strengths are shown without the contribution from the $^{61}$Cr--$^{61}$V Urca pair studied in this work (orange circle) and with the new data on $^{61}$Cr--$^{61}$V from this work (red error bar). Results shown are for a neutron star radius $R=12$~km and a temperature $T=0.5$~GK and will scale with $R^2T^5$ for different physical parameters.}
\label{CoolingStr}
\end{figure}

With our result we can now calculate the neutrino cooling from $A=61$ nuclei in neutron star crusts solely using experimental data\footnote{The experimentally deduced log $ft$ values for the ground-state to ground-state transitions in the other Urca pairs along the $A=61$ mass chain may have, unlike our measurement, significant additional systematic uncertainties from the Pandemonium effect. The cooling contribution estimated from these other Urca pairs is therefore an upper limit.}.  To determine crust cooling for accreting neutron stars that exhibit rp-process bursts, we folded the calculated cooling rates in individual mass chains with the rp-process ash abundances from \cite{Cyburt2016}. The result is shown in Fig.~\ref{CoolingStr}. Our deduced ground-state to ground-state log $ft$ value of 5.5$^{+0.2}_{-0.2}$ from this work is significantly higher than the theory value of 4.35 predicted for the lowest-lying transition by the QRPA-fY model and results in an Urca cooling rate slower by a factor of 14. It is consistent with the lower limit of 4.6 implied by the 40\% upper limit of the ground-state to ground-state branch from \cite{Suchyta2014}. Our most important conclusion is that our result clearly rules out the large cooling contribution from $A=61$ ashes predicted when employing the QRPA-fY $^{61}$V $\beta$-decay log $ft$ value for the transition to the 10~keV state in $^{61}$Cr. Nevertheless we find that neutrino cooling from $A=61$ is significantly larger than predicted when taking QRPA-fY ground-state to ground-state transitions at face value. 

We also find that based on our results the $^{61}$Cr--$^{61}$V Urca pair still makes the $A=61$ mass chain one of the strongest cooling chains in accreted neutron star crusts (Fig.~\ref{CoolingStr}). Only $A=31,33$, and 55 nuclei are predicted to provide stronger neutrino cooling than $A=61$.  Reliable experimental data are still lacking for ground-state to ground-state transitions within those other isobaric chains. We note it was recently pointed out that neutron transfer reactions may alter the distribution of abundances across mass chains from the initial burst ashes distribution \cite{Chugunov2019}. This may alter the relative weight of the individual mass chains in the ash composition. Taking this effect into account would require a significantly expanded reaction network with detailed sets of realistic ashes, which is under development but beyond the scope of this experimental paper. Our data on the $^{61}$Cr-$^{61}$V Urca pair will be an important input in such future calculations. 

In summary we have developed an experimental approach to infer $\beta$-decay ground-state to ground-state transition strengths for neutron-rich nuclei, and demonstrated the importance of such measurements to obtain reasonably accurate data on Urca cooling in accreted neutron star crusts. The Urca cooling from $A=61$ nuclei can now be determined based on experimental data. It will be important to also investigate the strong predicted cooling in the $A=31$, $A=33$, and $A=55$ mass chains. In particular $A=55$ is now the strongest predicted cooling mass chain within the mass range of the rp-process ashes.  All relevant transitions in these mass chains are within reach for future studies with the experimental approach developed in this work. 

We thank M. Emeric and A. Sonzogni for creating the LOGFT web tool. This work was conducted with the support of Michigan State University, the National Science Foundation under Grants PHY-1102511, PHY-1404442, PHY-1713857, PHY-1430152 (JINA Center for the Evolution of the Elements), and AST-1516969. It was additionally supported by the Department of Energy National Nuclear Security Administration through Award Numbers DE-NA-0003221 and DE-NA-0002132 and under the Nuclear Science and Security Consortium under Award Number(s) DE-NA0003180 and DE-NA0000979, and performed under the auspices of the U.S. Department of Energy by Lawrence Livermore National Laboratory under Contract DE-AC52-07NA27344. A. Spyrou would like to acknowledge support under NSF career grant PHY-1350234.

\bibliographystyle{apsrev4-1}
\bibliography{e14041paper}

\end{document}